\documentclass[a4paper,UKenglish]{lipics}
 
\usepackage{microtype}

\title{Using interrogative logic to teach classical logic\footnote{This work is part of the project on the interrogative model of inquiry and education currently under development at Yachay Tech.}}
\titlerunning{Using interrogative logic to teach classical logic} 

\author[1]{Levis Zerpa}
\affil[1]{Department of Social Science and Innovations, Yachay Tech\\
  Hacienda San José y Proyecto Yachay, Urcuquí, Ecuador\\
  \texttt{lzerpa@yachaytech.edu.ec}}
\authorrunning{L.\,Zerpa} 

\Copyright{Levis Ignacio Zerpa}


\serieslogo{logo_ttl}
\volumeinfo
  {M. Antonia {Huertas}, Jo\~ao {Marcos}, Mar\'ia {Manzano}, Sophie {Pinchinat}, \\
  Fran\c{c}ois {Schwarzentruber}}
  {5}
  {4th International Conference on Tools for Teaching Logic}
  {1}
  {1}
  {281}
\EventShortName{TTL2015}

\begin{document}

\maketitle

\begin{abstract}
In the paper I discuss a tool for helping students in their symbolizations of natural language sentences using the formal language of classical first order logic (CFOL). The tool is an extension of Hintikka's concept of (Inquirer’s) range of attention in the context of interrogative games. Any given text is reconstructed as the answer to a “big” or principal question obtained through the answers of a series of “small” or operative questions. The tool brings some “narrative flavor” to the symbolization and offers a convenient mold that can be used by students in many different contexts.
\end{abstract}

\section{Introduction}

Learning to symbolize natural language sentences in the formal language of classical first order logic (CFOL) is one of the main tasks of most logic courses. The concept of (Inquirer’s) range of attention RA (Hintikka \& Hintikka~\cite{HintikkaAndHintikka}, Genot \& Gulz~\cite{GenotAndGulz}, Zerpa~\cite{Zerpa}), from Hintikka’s interrogative logic, is a major tool to carry out that task. According to  Hintikka~\cite{Hintikka}, RA is the set of available tautological premises of the form $S \vee \neg S$. Intuitively speaking, the set RA codifies the totality of yes-or-no questions which the Inquirer is prepared to ask. If one considers an extension of the concept of range of attention according to which wh-questions, besides yes-or-no questions, are considered, then one is able to apply the RA concept as a tool to teach students how to use the expressions of the formal language of CFOL to symbolize natural language sentences. These expressions include sentence letters, truth-functional connectives, individual constants and variables, predicates, and quantifiers. The tool can be conceived as a procedure consisting of the following steps: first, reconstruct the text as the answer to a “big” or principal question obtained through the answers of a series of “small” or operative questions. Second, obtain the presupposition of each question dropping the question mark on it. Third, make the symbolization of all the previous questions and answers using the abovementioned expressions of the formal language of CFOL in the usual way. Fourth, make all the deductive inferences required. Fifth, organize the information previously obtained in a tableau like the ones described in Hintikka~\cite{Hintikka}. Yes-or-no questions, read as “Is it the case that $S$ or $\neg S$?”, have the logical form “$S \vee \neg S$?” while which-questions, read as “Which individual $x$ is such that $S(x)$?”, have the form “$\exists xS(x)$?”.  The presupposition of the yes-or-no question “$S \vee \neg S$?” is the tautology $S \vee \neg S$ while the presupposition of the which-question “$\exists xS(x)$?” is the sentence $\exists xS(x)$. The text under consideration may have a specific deductive structure or not; the point is that the strategy of break down the initial principal questions in a set of several operative questions of the forms specified brings a “narrative flavor” to the symbolization and, in that sense, constitutes a convenient mold that can be used by students in many different contexts. The logical justification of the procedure is found in theorem 7.49 of Wiśniewski~\cite{Wiśniewski} using multiple-conclusion entailment.

In the paper I report an ongoing research project related to the application of interrogative logic and the interrogative model of inquiry and learning to the teaching of elementary classical logic in introductory university courses. The empirical basis of the project is obtained from logic courses taught at Yachay Tech (Ecuador) during 2014-2015. In spite of the fact that the project is concerned with several aspects of the interrogative approach, in this paper I focus the attention on both the interrogative approach to the teaching of the material implication and the concept of range of attention as a tool to symbolize natural language sentences in CFOL.

The paper is organized as follows: first, some general remarks are made about interrogative logic and the interrogative model of inquiry and learning. Then, the interrogative approach to the teaching of material implication are briefly considered using the experience in Yachay Tech. Finally, the largest part of the paper is focused on the RA concept through the analysis of two concrete examples, one from everyday reasoning and other from scientific (experimental) reasoning. 

\section{Interrogative logic and the interrogative model of inquiry and learning}

Interrogative or erotetic logic is the logical theory of questions and answers. More specifically, interrogative logic studies formal systems in which interrogative as well as declarative sentences are formalized with precision. Following Frege~\cite{Frege}, questions are understood as requests for information and these requests are studied, in a detailed way, in these formal systems. Furthermore, it is known that questioning is a fundamental activity in knowledge-acquisition; it plays a crucial role in both scientific research and the learning process. This idea leads toward the interrogative approach to inquiry and learning: both inquiry and learning are processes which can be modeled by means of sequences of questions and answers. More specifically, progress in inquiry can be represented through a study of the transformation of the successive question asked by these agents and the formulation of answers to them. According to Hintikka´s Interrogative Model of Inquiry the process of inquiry is modeled as a two-person game between an inquirer and a source of information. The goal of the game is to get a conclusive answer to the principal (or “big”) question posed by the inquirer through the formulation of more specific (or “smaller”) questions called operative questions. The moves of the game are of two types: interrogative and deductive. Interrogative moves consist of questions or answers while deductive moves are deductive inferences obtained from previous moves. The score of the game is kept in semantic tableaus (Hintikka~\cite{Hintikka}). The Interrogative Model of Learning is an application of the interrogative model of inquiry to the study of educational practices emphasizing the importance of the transformation of questions through the learning process (Hakkarainen \& Sintonen~\cite{HakkarainenAndSintonen}).

\section{Teaching material implication through questioning}

Traditionally, material implication is a perplexing topic for many students at the very beginning of their learning process. The truth table of the material implication, specially the cases in which the antecedent are false (the FV and FF cases in which, according to material implication, the resulting conditional is true), are usually a source of confusion and perplexity for many students. After a study of negation, conjunction and disjunction, students were exposed with four examples of  conditional statements of a very diverse nature, and the principal question asked to them was “when all these conditional are false?”. Students provided two responses: (1) the conditional is false when VF is the case and (2) when FF is the case. On the basis of an ilustrative example, operative questions, answers to them, and further information, they discarded (2) in favor of (1). Furthermore, the symbolization of arguments in natural language containing negation, conjunction, disjunction and material implication were guided by the processing of questions. Given an argumentative text, the principal question, “which is the correct symbolic translation of this text?”, was transformed into a first “intermediate” operative question: “which are the indicator words in the text?”. This last questions was transformed into three “smaller” (= more specific) operative questions: “which are the conclusion indicators?”, “which are the premise indicators?”, “which are the phrases, in the text, related to truth-functional connectives?”. In turn, this last operative question was transformed into four more specific operative questions of the form “which phrase of the text can be translated with an $X$?” where $X$ stands for a truth-functional connective (the biconditional was introduced shortly after).
Even specific logical laws were directly approached from an interrogative viewpoint, For example, the De Morgan theorems were presented as the answer to the question “how one can distribute the negation in a formula of the form "$\neg(A \vee B)$". My hypothesis is this: the main logical laws can be taught as the correct answers to questions about inference. One can motivate students to ask these questions so they can discover, by themselves, these laws as their answers to their own questions. A major tool in that process is the transformation of their questions. This approach is closely related to Bereiter~\cite{Bereiter} problem-solving approach to learning. 
A precise framework to study the transformation of questions is the abovementioned concept of range of attention. In this framework students are able to use the expressions of the formal language of CFOL to symbolize texts in natural language. Students start by consider the available questions in the RA set. Then, they organize them in such a way that the structure of the text is conceived as the answer to a principal question by means of a series of operative questions.

\section{The RA set and its use in the teaching of predicate logic (1): Holmes’ reasoning}

Arthur Conan Doyle’s story Silver Blaze can be summarized in the following way: the horse Silver Blaze has been stolen and its owner found dead. To solve the case, Holmes must find out who stole the horse. During the investigation, the following exchange takes place (Hintikka~\cite{Hintikka}, Genot \& Gulz~\cite{GenotAndGulz}):

\emph{Watson: Is there any point to which you would wish to draw my attention?}

\emph{Holmes: To the curious incident of the dog in the night-time.}

\emph{Watson: The dog did nothing in the night-time.}

\emph{Holmes: That was the curious incident.}

There is a remarkable step of reasoning here, namely, the identification of the violation of a natural expectation: a dog would have barked on a stranger. But it did not. So, the thief was known by the dog. Furthermore, it is determined that the only relevant individuals are the following: the unknown thief, the dead stable-master (the owner), and his killer. Therefore, from the reduced number of individuals and the curious incident of the dog in the night-time, Holmes correctly infers that the thief is the owner. This piece of reasoning is reconstructed by means of the concept of Inquirer’s range of attention, taking Holmes as the Inquirer. According to the intuitive interpretation of the set RA, it codifies the totality of yes-or-no questions which Holmes is prepared to ask. In this case, a small proper subset RA$_H$ of RA is enough. Following a sequential order, the two main (operative) yes-or-no questions in RA$_H$ are the following: “Is there a dog in the stable or not?” (answer: yes) and “did the dog barked at the thief or not?” (answer: no). According to the extension of the RA concept I am proposing, in RA one is able to ask wh-questions besides yes-or-no questions. Therefore, the principal question “who is the thief?” can be included in RA$_H$. Consequently, if constants o and t stand for the owner and the thief, respectively, the tableau of the game starts with the principal question “who is the thief?” or “$\exists (x = t)$?” and ends with the conclusive answer to this question, namely, “the thief is the owner” or “$t = o$”. Holmes’ reasoning is reconstructed as an interrogative game which is played through a series of yes-no questions from RA$_H$, their answers, and deductive inferences from them. In general, a yes-no question of the form “$S \vee \neg S$?” has presupposition $S \vee \neg S$, so the question only presupposes the truth of that tautology. And a known theorem of interrogative logic (see Hintikka~\cite{Hintikka}) guarantees that a question like this always can be properly asked (under the appropriate conditions). The sequence of questions and answers are the following (all the interrogative moves are carried out by the inquirer, Holmes, and the answers are provided by the source of information):

(1)	\textbf{Interrogative move}: \emph{Is there a dog in the stable or not?} In symbols,
 
“$\exists x$[dog\textunderscore in \textunderscore the \textunderscore stables($x$)] $\vee \neg \exists x [$dog\textunderscore in \textunderscore the \textunderscore stables($x$)]?”

(2)	\textbf{Answer}: \emph{yes}; in symbols, $\exists x[$ dog\textunderscore in \textunderscore the \textunderscore stables($x$)].

(3)	\textbf{Deductive inference move}: dog\textunderscore in \textunderscore the \textunderscore stables($d$) by existential instantiation ($d$ is a new constant). This step is similar to say “let’s call $d$ the dog in the story”.

(4)	\textbf{Interrogative move}: \emph{Did d barked at the thief or not?} In symbols, 
$d$ barked at $t \vee \neg(d$ barked at $t$)? 

(5)	\textbf{Answer}: \emph{no}; in symbols, $\neg(d$ barked at $t$).

(6)	\textbf{Interrogative move (non-triviality of Holmes’ reasoning)}: Holmes then asks whether the dog did not bark to a specific person or not. In other words, the dog did not bark to the thief but it would have barked to anybody else. This question rules out the possibility that the dog just did not bark to anybody (for example, if the dog was sent to sleep by somebody), so this move avoids the trivialization of Holmes’ reasoning. In a semi-symbolic way, the question is this: “is the individual to whom the dog didn’t bark unique or not?”. In symbols, 
$\exists z \forall y [\neg (d$ barks at $y$) $\rightarrow y = z] \vee \neg \exists z \forall y [\neg (d$ barks at $y$) $\rightarrow y = z]$?

(7)	\textbf{Answer}: \emph{yes}; in symbols, $\exists z \forall y [\neg (d$ barks at $y$) $\rightarrow y = z]$.

(8)	\textbf{Deductive inference move}: By existential instantiation, $(d$ barks at $t) \rightarrow t = o$.

(9)	\textbf{Deductive inference move}: By modus ponens from answer (5) and deductive inference move (8), the case is solved: $t = o$, the thief is the owner.

In the corresponding tableau, the presupposition of any question must be established before that question can asked. A nice feature of Hintikka’s theory is that the presupposition of any question is obtained simply by dropping the question mark on it. The traffic from side to side of the tableau also has restrictions. Some abbreviations are used: "$D(x)$" abbreviates "$dog\textunderscore in \textunderscore the \textunderscore stables(x)$", "$B(x, t)$" abbreviates "$x$ barked at $t$", and "$unique$" abbreviates "`$\exists z \forall y [\neg (d$ barks at $y$) $\rightarrow y = z]$". In summary, the tableau looks like this: 

\begin{table*}[htbp]
\begin{tabular}{cccc}
 
 \\
 $\#$&SOURCE OF INFORMATION&INQUIRER: HOLMES&$\#$\\ \hline
 1 & $\exists x (x = t)$ & $\exists x (x = t)? $ & 2 \\ \hline
 3 & $\exists xD(x) \vee \neg \exists xD(x)$ & $\exists xD(x) \vee \neg \exists xD(x)$? & 4\\ \hline
 5 & $\exists xD(x)$ & $\exists xD(x)$? & 6\\ \hline
 7 & $D(d)$ &  & 8\\ \hline
 9 & $B(d, t) \vee \neg B(d, t)$ & $B(d, t) \vee \neg B(d, t)$? & 10\\ \hline
 11 & $\neg B(d, t)$ & & 12 \\ \hline
 13 & $unique \vee \neg unique$ & $unique \vee \neg unique$? & 14 \\ \hline
 15 & $unique $ & & 16 \\ \hline 
 17 & $\neg B(d, y) \rightarrow t = o $ &  & 18 \\ \hline
 19 & $ t = o$ (case solved!) &  & 20 \\ \hline
 \\
 \end{tabular}
\end{table*}

In this case, the basic steps of reasoning were obtained through yes-no questions. Only the principal question is a wh-question, namely, “who is the thief?”. But in more complex cases, wh-questions are needed through all the game. Mendel’s research on genetics can also be modeled as an interrogative game. The complete tableau is more complex than the previous one; see Zerpa~\cite{Zerpa} for details. The next subsection has a small subtableau representing the results of the first round of Mendel’s experiments on peas (\textit{pisum sativum}). 

\section{The RA set and its use in the teaching of predicate logic (2): Mendel’s experimental results}

Mendel´s principal question on the subgame considered is this: “Which are the results of the first experiment?”. To answer this principal question, Mendel asks to the source of information several operative questions that provide the answer to that principal question. In this case, the members of the set RA include which-questions about the number of plants crossed and the form and number of differentiating character selected. Mendel´s most innovative questions are (*) and (**). In the dialogical framework of the game, these which-questions are the following:

\textbf{Mendel}: \emph{Which is the number of plants crossed?} 

\textbf{Source of information}: \emph{253 plants.} 

\textbf{Mendel}: \emph{Which is the (constantly) differentiating character selected?} 

\textbf{Source of information}: \emph{form of the seed.} 

\textbf{Mendel}: \emph{Which character is dominant?} 

\textbf{Source of information}: \emph{round (seed).}

\textbf{Mendel}: \emph{Which character is recessive?} 

\textbf{Source of information}: \emph{wrinkled (seed).}

\textbf{Mendel}: \emph{Which are the numbers of different forms under which the progeny appeared in the first experiment?} 

[This question is transformed into two operative questions:]

\textbf{Mendel}: \emph{Which is the number of plants showing the dominant character? (*)} 

\textbf{Source of information}: \emph{5,474 plants showed round seed.}

\textbf{Mendel}: \emph{Which is the number of plants showing the recessive character? (*)} 

\textbf{Source of information}: \emph{1,850 plants showed wrinkled seed.}

\textbf{Mendel}: \emph{Which are the ratios found between the hybrid forms in the first experiment? (**)} 

\textbf{Source of information}: \emph{A ratio of 2.96 : 1 between dominant and recessive characters appeared among the progeny in the first experiment.}

The entire tableau is large; to get an idea of it is convenient to consider the questions concerning the number of plants showing the dominant and recessive characters. Let’s introduce the following constants and predicates:
D1(x): x is a plant which shows dominant character D1 (in experiment 1); D1 = round seed,
R1(x): x is a plant which shows recessive character R1 (in experiment 1); R1 = wrinkled seed
N(x, y): x is the number assigned to y (after counting them.

Then, this part of the tableau looks like this:
	
	\begin{table*}[htbp]
\begin{tabular}{cccc}
 
 \\
 $\#$&SOURCE OF INFORMATION&INQUIRER: MENDEL&$\#$\\ \hline
 ... & ... & ... & ... \\ \hline
 1 & $\exists x \exists y[D_1 (x) \wedge N (y, x)]$ & $\exists x \exists y[D_1 (x) \wedge N (y, x)]$?(*) & 2 \\ \hline
 3 & $D_1 (a) \wedge N (5,474, a)$ & & 4\\ \hline
 5 & $\exists x \exists y[R_1 (x) \wedge N (y, x)]$ & $\exists x \exists y[R_1 (x) \wedge N (y, x)]$?(*) & 6\\ \hline
 7 & $R_1 (a) \wedge N (1,850, b)$ & & 8\\ \hline
 ... & ... & ... & ...\\ \hline
 \\
 \end{tabular}
\end{table*}

\section{Conclusion}

In conclusion, the concept of RA provides a precise framework to teach symbolization of texts in natural language through a specification of the transformation of a principal question in a series of operational questions. The resulting analysis keeps a certain “narrative flavor” that is sometimes absent from the usual teaching approaches. Furthermore, the interrogative model of learning provides an interesting way to motivates active learning through a study of the transformation of questions in logic courses.









\subparagraph*{Acknowledgements}

My thanks go to two anonymous referees for their helpful comments on the earlier version of this paper.





\end{document}